\documentclass[12pt]{article}  
\newcommand{\comma}{\;\; ,}
\newcommand{\period}{\;\; .}
\newcommand{\eq}{\; = \;}
\newcommand{\sep}{\;\; , \;\;}
\newcommand{\be}{\begin{equation}}
\newcommand{\bd}{\begin{displaymath}}
\newcommand{\ee}{\end{equation}}
\newcommand{\ed}{\end{displaymath}}
\newcommand{\ba}{\begin{eqnarray}}
\newcommand{\ea}{\end{eqnarray}}
\newcommand{\Wb}{\overline{W}}
\newcommand{\pb}{\overline{p}}
\newcommand{\qb}{\overline{q}}

\renewcommand{\i}{{\rm i}}

\newcommand{\inn}{\; {\scriptstyle \in } \;}

\newcounter{storeeqn}

\title{Derivation of the order parameter of the chiral 
Potts model}

\author{ R.J. Baxter\\
{\protect \small  Mathematical
Sciences Institute}\\
{\protect \small  The Australian National University,
 Canberra, A.C.T. 0200, Australia  }
}

\date{}

\begin{document}

%\magnification = \magstep1
%\magnification = 1000

\maketitle

{\footnotesize {\bf Abstract} We derive the order parameter of the 
chiral Potts  model, using the method of Jimbo {\it et al}. The result
agrees with previous conjectures.}

\vspace{5 mm}


\vspace{5 mm}

The solvable chiral Potts model is one of  $N$-state spins
with nearest-neighbour interactions on a planar lattice 
$\cal L$.\cite{BPAuY88} It is like other solvable models in that
its interactions  satisfy the star-triangle relations. However, 
it is unlike most of them in that it does not have the ``rapidity 
difference'' property. This makes the model mathematically much more 
difficult. The free energy of the model has been obtained using
functional transfer matrix relations,\cite{BBP90, RJB90, RJB91} but
the order parameters (spontaneous magnetizations) have so far 
defied calculation. In particular, the corner transfer matrix method
appears to fail completely.\cite{RJB93}

This is despite the fact that there is an elegant and eminently 
believable conjecture for these order 
parameters.\cite{HKN83,HK89,AMPT89}

Here we derive the order parameters and verify the conjecture. The 
method is quite similar to the ``inversion relation'' method 
for calculating the free energy.\cite{RJB03} Both involve assuming
certain analyticity properties of a ``$\tau_2(t_q)$'' model
(which is closely related to the superintegrable chiral Potts model).

%23456789012345678901234567890123456789012345678901234567890123456789012

We take $\cal L$ to be the square lattice,
drawn diagonally as in Fig. \ref{Fpqfig}.
Each spin $i\, $ takes one of $N$ possible states, 
labelled $0, \ldots , N-1$. It interacts with 
a neighbouring site $j$  with Boltzmann weight
$W_{vh}(\sigma_i - \sigma_j)$ or $\Wb_{vh}(\sigma_i - \sigma_j)$
if $i$ is below $j$ and the edge is in the SW-NE or SE-NW direction,
respectively.

Here $v, h$ each denote a set of rapidity variables. Let
$p = \{ x_p, y_p, \mu_p, t_p \}$, where $x_p, y_p, \mu_p, t_p$
are related to one another by
\be \label{xymu} x_p^N+y_p^N = k(1+x_p^N y_p^N) 
\sep k x_p^N = 1-k'/\mu_p^N 
\sep k y_p^N = 1-k'\mu_p^N \comma \ee
\bd  k^2 t_p^N \eq  (1-k'\mu_p^N)(1-k'/\mu_p^N)  \ed
and
\be k \eq (1-{k'}^2)^{1/2} \sep 0 < k,k' < 1  \period \ee
Here $k, k'$ are constants, the same at all sites of the lattice, and 
$k'$ is  a ``temperature-like'' variable, being small at low 
temperatures. Here we take $0 < k' < 1$, which means that the 
system is ferromagnetically ordered. It becomes 
critical when $k' \rightarrow 1$.

Rapidities such as  $p, q, v, h$ are associated with the dotted 
lines of Fig. \ref{Fpqfig} and may vary from line to line.
If the edges $(i,j)$, $(i,k)$ of $\cal L$  are intersected by  
rapidity lines $v,h$, ordered as in  Fig. \ref{Fpqfig}, then the edge 
weight functions
are $W_{vh}(\sigma_i - \sigma_j)$, $\Wb_{vh}(\sigma_i - \sigma_k)$ 
where
\ba \label{WWba}
W_{pq}(n) & = & (\mu_p/\mu_q)^n \prod_{j=1}^n \frac{y_q - \omega^j x_p}
{y_p - \omega^j x_q} \comma  \\
\label{WWbb}
\Wb_{pq}(n) & = &  (\mu_p \mu_q)^n  \prod_{j=1}^n \frac{\omega x_p -   
\omega^j x_q} {y_q - \omega^j y_p} \comma \ea
where
$\omega \eq \exp (2\pi \i/N) $
and $v = \{ x_v, y_v, \mu_v, t_v \}$, 
$h = \{ x_h, y_h, \mu_h, t_h \}$ are rapidity sets satisfying
(\ref{xymu}).

If  $x_p,x_q,y_p,y_q,\omega x_p$ all lie on the unit circle 
in an anti-clockwise sequence, then $W_{pq}(n)$, 
$\Wb_{pq}(n)$ are real and positive. We refer to this as 
the physical case.

Hold the spin $a$ in  Fig. \ref{Fpqfig} fixed.
Then the partition function is 
\be \label{pfn}
Z(a)  \eq \sum_{\sigma} \, \prod W_{vh}(i-j)  \prod 
\Wb_{vh}(j-k) \comma \ee
the products being over all edges of each type
(with their appropriate rapidity variables), and the sum over
all values of the other spins, the boundary spins being set to zero. 
The unrestricted partition function is
$Z = Z(0) + \cdots + Z(N-1)$.

%%%%%%%%%  PICTURE OF F_{pq} %%%%%%%%

%% The following picture is allowed a space 
\setlength{\unitlength}{1pt}
\begin{figure}[hbt]
\begin{picture}(420,260) (0,0)

\multiput(30,15)(5,0){73}{.}
\multiput(30,75)(5,0){32}{\bf .}
\multiput(202,75)(5,0){35}{\bf .}
\multiput(29,75)(5,0){32}{\bf .}
\multiput(203,75)(5,0){35}{\bf .}
\multiput(30,135)(5,0){73}{.}
\multiput(30,195)(5,0){73}{.}
\put (190,72) {\line(0,1) {8}}
\put (200,72) {\line(0,1) {8}}
\thicklines

\put (69,72) {\large $< $}
\put (70,72) {\large $< $}
\put (71,72) {\large $< $}

\put (308,12) {\large $< $}
\put (309,12) {\large $< $}
\put (310,12) {\large $< $}

\put (308,72) {\large $< $}
\put (309,72) {\large $< $}
\put (310,72) {\large $< $}

\put (308,132) {\large $< $}
\put (309,132) {\large $< $}
\put (310,132) {\large $< $}

\put (308,192) {\large $< $}
\put (309,192) {\large $< $}
\put (310,192) {\large $< $}

\put (42,230) {\large $\wedge$}
\put (42,229) {\large $\wedge$}
\put (42,228) {\large $\wedge$}

\put (102,230) {\large $\wedge$}
\put (102,229) {\large $\wedge$}
\put (102,228) {\large $\wedge$}

\put (162,230) {\large $\wedge$}
\put (162,229) {\large $\wedge$}
\put (162,228) {\large $\wedge$}

\put (222,230) {\large $\wedge$}
\put (222,229) {\large $\wedge$}
\put (222,228) {\large $\wedge$}

\put (342,230) {\large $\wedge$}
\put (342,229) {\large $\wedge$}
\put (342,228) {\large $\wedge$}

\thinlines

%%%%%% FIRST ONE %%%%%%%%

\put (176,102) {{\Large \it a}}
\put (320,60) {{\Large \it q}}
\put (83,60) {{\Large \it p}}
\put (380,-2) {{\Large \it h}}
\put (380,118) {{\Large \it h}}
\put (380,178) {{\Large \it h}}

\put (70,115) {{\Large \it i}}

\put (130,145) {{\Large \it j}}
\put (10,145) {{\Large \it k}}

\put (195,105) {\circle{7}}

\put (16,45) {\line(1,-1) {60}}
\put (16,165) {\line(1,-1) {180}}
\put (76,225) {\line(1,-1) {117}}
\put (198,103) {\line(1,-1) {117}}
\put (196,225) {\line(1,-1) {180}}
\put (316,225) {\line(1,-1) {60}}
\put (16,165) {\line(1,1) {60}}
\put (16,45) {\line(1,1) {180}}
\put (76,-15) {\line(1,1) {117}}
\put (198,107) {\line(1,1) {118}}
\put (196,-15) {\line(1,1) {180}}
\put (316,-15) {\line(1,1) {60}}

\put (75,105) {\circle*{7}}
\put (315,105) {\circle*{7}}
\put (75,-15) {\circle*{7}}
\put (195,-15) {\circle*{7}}
\put (315,-15) {\circle*{7}}

\put (15,45) {\circle*{7}}
\put (135,45) {\circle*{7}}
\put (255,45) {\circle*{7}}
\put (375,45) {\circle*{7}}

\put (15,165) {\circle*{7}}
\put (135,165) {\circle*{7}}
\put (255,165) {\circle*{7}}
\put (375,165) {\circle*{7}}

\put (75,225) {\circle*{7}}
\put (195,225) {\circle*{7}}
\put (315,225) {\circle*{7}}

\put (42,-40) {{\Large \it v}}
\put (102,-40) {{\Large \it v}}
\put (162,-40) {{\Large \it v}}
\put (222,-40) {{\Large \it v}}
\put (282,-40) {{\Large \it v}}
\put (342,-40) {{\Large \it v}}

\multiput(45,-25)(0,5){52}{.}
\multiput(105,-25)(0,5){52}{.}
\multiput(165,-25)(0,5){52}{.}
\multiput(225,-25)(0,5){52}{.}
\multiput(285,-25)(0,5){52}{.}
\multiput(345,-25)(0,5){52}{.}
 \end{picture}
\vspace{1.5cm}
\caption{ The square lattice with the cut rapidity line below $a$.}
\label{Fpqfig}
\end{figure}

We have followed Jimbo {\it et al}~\cite{JMN93} and cut 
the horizontal rapidity line immediately  below $a$, giving the 
left (right) half-line a rapidity $p$ ($q$). Define
\be F_{pq}(a) \eq Z(a)/Z \period \ee
We expect this ratio to tend to a limit when the lattice is 
large. It will {\em not} depend on the ``background'' rapidities
$v,h$ because of the star-triangle relation,\cite{BPAuY88} which
allows us to  move any of the $v, h$ rapidity lines infinitely 
far away from  the spin $a$.

However, we cannot move the half-lines $p, q$ away from
$a$, so $F_{pq}(a)$ will indeed be a function of $p,q$. It is the 
probability that the central spin has value $a$. An exceptional case 
is  when
$q = p$, when the cut disappears and the recombined line can be moved 
to infinity, so
\be \label{Fpp}
F_{pp}(a) \eq {\;  \rm independent \; \;  of \; \; } p 
\period \ee

What we can do is rotate $p,q$ round $a$ and then interchange them, 
which  gives us some functional relations satisfied by 
$F_{pq}(a)$.\cite{RJB98}
If we  define
\be \label{fourF}
\tilde{F}_{pq}(r) \eq \sum_{a=0}^{N-1} \omega^{r a} 
\, F_{pq}(a) \comma \ee
\be \label{defGpq}
G_{pq}(r) \eq \tilde{F}_{pq}(r) /\tilde{F}_{pq}(r-1)  \comma \ee
then for the infinite lattice we obtain the relations
\addtocounter{equation}{1}
\setcounter{storeeqn}{\value{equation}}
\setcounter{equation}{0}
\renewcommand{\theequation}{\arabic{storeeqn}\alph{equation}}

\be \label{10a}
G_{Rp,Rq}(r) \eq 1/G_{pq}(N-r+1) \comma \ee
\be \label{10b}
G_{pq}(r) \eq  G_{Rq, R^{-1} p}(r) \comma \ee
\be \label{10c}
G_{pq}(r) \eq  \frac{x_q \mu_q - x_p \mu_p \, \omega^r}
{y_p \mu_q - y_q \mu_p \, \omega^{r-1}} \; G_{R^{-1}q, R p}(r) 
 \comma \ee
\be \label{10d}
\prod_{r=1}^N G_{pq}(r) \eq 1 \comma \ee
\be \label{10e}
G_{Mp,q}(r) \eq G_{p,M^{-1} q}(r) \eq G_{pq}(r+1) \period \ee

\setcounter{equation}{\value{storeeqn}}
\renewcommand{\theequation}{\arabic{equation}}

Here $R$, $M$ are automorphisms that act on the rapidities:
\be \label{defR}
\{ x_{Rp}, y_{Rp}, \mu_{Rp}, t_{Rp} \} \eq 
\{ y_p, \omega x_p, 1/\mu_p, \omega  t_p \} \period  \ee

\be \label{defM}
\{ x_{Mp}, y_{Mp}, \mu_{Mp}, t_{Mp} \} \eq 
\{ x_p, y_p, \omega\mu_p,  t_p \} \period  \ee

We regard $t_p$ as an independent complex variable and
$x_p, y_p, \mu_p$ as determined from it
by (\ref{xymu}). They are multi-valued
functions of $t_p$: to make them single-valued we must cut the 
$t_p$-plane as in Fig. \ref{brcuts}. There are $N$ 
cuts ${\cal C}_0, \ldots , {\cal C}_{N-1}$,
where ${\cal C}_j$ lies on the radial line $\arg(t_p) = 2 \pi j/N$.

%%%%%%%%%  Branch cuts in the complex $t_p$-plane %%%%%%%%

%% The following picture is allowed a space 

\setlength{\unitlength}{1pt}
\begin{figure}[hbt]
\begin{picture}(420,260) (0,0)

\put (50,125) {\line(1,0) {350}}
\put (225,0) {\line(0,1) {250}}

\put (325,125)  {\circle*{9}}
\put (175,208)  {\circle*{9}}
\put (175,42)  {\circle*{9}}

\put (315,100)  {\Large 1}
\put (185,214)  {\Large $\omega$}
\put (184,32)  {\Large $\omega^{2}$}

\put (338,140) {{\Large {${\cal C}_0$}}}

\put (163,176) {{\Large {${\cal C}_1$}}}

\put (145,37) {{\Large {${\cal C}_2$}}}

\put (305,10) {{\Large {$t_p$-plane}}}
\thicklines
\put (295,124) {\line(1,0) {60}}
\put (295,125) {\line(1,0) {60}}
\put (295,126) {\line(1,0) {60}}

\put (160,16) {\line(3,5) {30}}
\put (160,17) {\line(3,5) {30}}
\put (160,18) {\line(3,5) {30}}

\put (160,234) {\line(3,-5) {30}}
\put (160,233) {\line(3,-5) {30}}
\put (160,232) {\line(3,-5) {30}}

\thinlines

\end{picture}
\vspace{1.5cm}
\caption{ The $t_p$-plane for $N=3$, showing the cuts
${\cal C}_0$, ${\cal C}_1$, ${\cal C}_2$.}
\label{brcuts}
\end{figure}
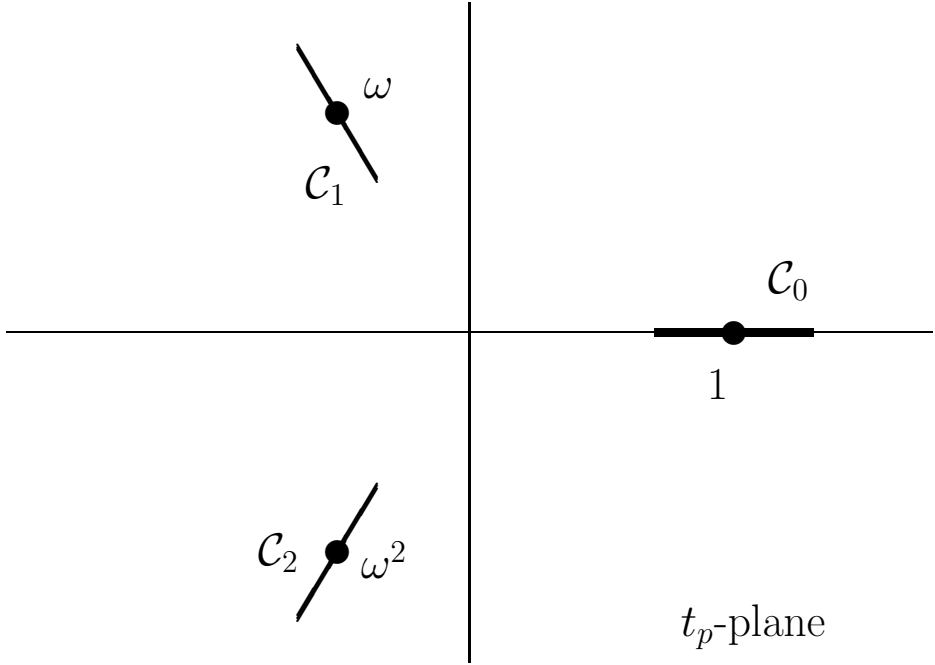

 The case we shall be interested
in is when $|\mu_p |>1$ and $\arg (x_p)$ is between 
$-\pi/(2N)$ and $\pi/(2 N)$. Then  $x_p$ lies in a small approximately 
circular region ${\cal R}_0$ round $x_p = 1$, while $y_p$ lies in 
a plane with $N$ corresponding approximately circular holes 
${\cal R}_0, \ldots , {\cal R}_{N-1}$ surrounding the points
$1, \omega, \ldots, \omega^{N-1}$. We say that $p$ lies in the
domain ${\cal D}_1$.

%23456789012345678901234567890123456789012345678901234567890123456789012

In the low temperatures  limit  $k' \rightarrow 0 $  and  
${\cal C}_j$, ${\cal R}_j$  both shrink to the 
point $\omega^j$. Then $x_p  \rightarrow 1 $ and $y_p$ can lie anywhere 
except at a root of unity.  We take $p, q, h \inn {\cal D}_1$ and 
$\mu_v = {\rm O}(1)$, $x_v \simeq y_v \simeq 1$. Then $W_{vh}(n),
\Wb_{vh}(n)$ are small unless $n = 0$ (mod $N$), which is the usual 
low-temperature case. The sum in (\ref{pfn}) is dominated
by the contribution from all spins other than $a$ being zero, and we 
obtain
\be \label{Fpqlt}
F_{pq} (a) \eq \frac{{k'}^2 (\mu_p/\mu_q)^a }{N^2 
(1-\omega^a)(1-\omega^{-a}) }\; \prod_{j=1}^a \frac 
{1-\omega^{j-1} t_q} {1-\omega^{j-1} t_p} \period \ee

Let us take $q$ to be related to $p$ by
\be \label{spcase}
x_q = x_p \sep y_q = \omega y_p \sep \mu_q = \mu_p \period \ee
Then (\ref{Fpqlt}) simplifies to 
\be \label{Fpqlt2}
F_{pq} (a) \eq \frac{{k'}^2  }{N^2 
(1-\omega^a)(1-\omega^{-a}) } \; \frac {1 -\omega^{a} t_p}
 {1- t_p}   + {\rm O} ({k'}^4) \period \ee
Thus to leading order in $k'$ we see that  $F_{pq} (a)$ is an
analytic and bounded function of $t_p$ except at $t_p=1$. This is 
consistent with $F_{pq} (a)$ and $G_{pq}(r)$  having branch cuts 
(for non-zero $k'$) on ${\cal C}_0$, but there is no indication of 
branch cuts at   ${\cal C}_1, \ldots , {\cal C}_{N-1}$.

We argue that this is exactly true for sufficiently small non-zero 
values of  $k'$.  If we 
rotate and reverse the left half-line $p$ anticlockwise below $a$,
as in \cite{RJB98}, it becomes a left-pointing line with
rapidity $p' = R^{-1} p$.\footnote{We can then interchange 
$p'$ with $q$, 
which gives the symmetry (\ref{10b}).} (To ensure that at low 
temperatures the dominant contribution continues to come from all 
non-central spins being zero, we should reduce 
$\mu_v$ to order $k'$, $\mu_h$ to order unity, and take
$x_h \simeq y_h \simeq 1$.)

The half-line $p'$ now lies immediately below $q$, and {from} 
(\ref{spcase}) 
\be 
x_q = y_{p'} \sep y_q = \omega^2 x_{p'} \sep 
 \mu_q = 1/ \mu_{p'} \period \ee
This is precisely the condition for the combined weights
of the two half-rows $p', q$ (summed over intervening spins)
to be those of the $\tau_2(t_{p'})$ model. (Eqns. 2.38 to 3.48 of
\cite{BBP90}, with $k = 0$, $\ell = 2$.). For a finite lattice, if
there are $s$ spins to the left of $a$, 
then  $Z(a)$ depends on $p$ only via $t_p$, and is a 
{\em polynomial } of degree $s$ in $t_p$.

Further, when $k'$ is small and $a = 0$, this polynomial 
is $(1-t_p)^s$, and is small (of order ${k'}^2$) when $0 < a < N$.
Hence $G_{pq}(r)$ is the ratio of two polynomials, each of degree 
$s$ and tending to $(1-t_p)^s$ as $k' \rightarrow 0$.

By continuity, for sufficiently small but non-zero $k'$
the zeros of the polynomials must be close to $t_p = 1$,
i.e. $t_{p'} = 1/\omega$.

This is the same behaviour as the free energy of the 
$\tau_2(t_{p'})$ model. This leads us
{\em assume}, corresponding to  Assumption 2 of \cite{RJB03},
that $G_{pq}(r)$ is analytic in the cut $t_p$ plane of
Fig. \ref{brcuts}, except only for the branch cut 
${\cal C}_0$.\footnote{One can obtain additional evidence for 
the analyticity  near ${\cal C}_0$ from the figure 
obtained by rotating $p$ clockwise to become a half-line
$Rp$ lying above $q$.\cite{RJB05}} 

Now look at (\ref{10c}). Using (\ref{spcase}) it becomes,
for $r = 1, \ldots , N-1$, 
\be \label{nobrcut}
x_ p^{-1} G_{pq}(r) \eq  x_{\pb}^{-1} G_{\pb, \qb}(r) \comma \ee
where $\pb = R^{-1}q, \qb = Rp$. It follows that $\pb, \qb$ satisfy
the relation (\ref{spcase}) and $x_{\pb} = y_p, y_{\pb}  = x_p,
\mu_{\pb} = 1/\mu_p$. The LHS of this equation is therefore the same 
as the RHS, with the same value of $t_p$, but with $x_p, y_p$ 
interchanged and $\mu_p$ inverted.

This is what happens if one crosses the branch cut ${\cal C}_0$
in Fig. \ref{brcuts} and then returns to the original value of $t_p$.
Thus (\ref{nobrcut}) is equivalent to the statement that
$x_ p^{-1} G_{pq}(r) $ does {\em not} have a branch cut at 
${\cal C}_0$.

Write $G_{pq}(r)$ as $G_r (t_p)$ and consider the function
\be L_r(t_p) \eq x_p^{-1 } G_r (t_p) G_r (\omega t_p) \cdots 
G_r (\omega^{N-1}t_p) \period \ee
For $r \neq 0$ the factor $x_p^{-1 } G_r (t_p) $ has no cut 
on ${\cal C}_0$.
{From} our assumption, neither do any of the other $G$ factors.
Hence $ L_r(t_p)$ has no such cut. It is unchanged by 
$t_p \rightarrow \omega t_p$, so has no cuts at any of the 
 ${\cal C}_j$. When $t_p, y_p, \mu_p \rightarrow \infty$ (their 
ratios remaining finite and non-zero) the Boltzmann
weights remain finite, so we expect $ L_r(t_p)$ to be bounded at 
infinity. From Liouville's theorem it is therefore a constant, 
so
\be L_r(t_p) \eq C_r \sep   0 < r < N \comma \ee
$C_r$ being a constant.\footnote{For $N=3$ we first observed this
{from} our previous series expansions.\cite{RJB98b}}

When $t_p, y_p  = 0$, then $q=p$ and $x_p^N = k$. If 
$t_p, y_p, \mu_p \rightarrow \infty $, then $x_p^N = 1/k$
and $q = M^{-1} p$. From (\ref{Fpp}) and (\ref{10e})
it follows that
\be C_r =  k^{-1/N} G_{pp}(r)^N = k^{1/N} \, G_{pp}(r+1)^N \comma \ee
for $r = 1, \ldots , N-1$.

Eliminating $C_r$ and using (\ref{10d}), we obtain
\be G_{pp}(r) = k^{(N+1-2r)/N^2}  \ee
for $r = 1, \ldots , N$.

The order parameter of the chiral Potts model is 
\ba \langle \omega^{r a} \rangle & = & \tilde{F}_{pq}(r)/ 
\tilde{F}_{pq}(0) \nonumber \\
 & = & G_{pp}(1) \cdots G_{pp}(r) \comma \ea
so we have 
\be 
\langle \omega^{r a} \rangle = k^{r (N-r)/N^2}  \ee
for $r = 0, \ldots, N$. This is the result previously conjectured on 
the basis of series  expansions (eqn. 1.20 of 
Ref. \cite{AMPT89}).  We expect all functions to be analytic in the 
physical ferromagnetically ordered regime (with positive real 
Boltzmann weights) so our working and results should remain true  
throughout $0 < k' < 1$. The magnetic critical exponents are 
$\beta_r = r (N-r)/(2 N^2)$.

%23456789012345678901234567890123456789012345678901234567890123456789012 

In \cite{RJB05} we go on to obtain $G_{pq}(r)$ from a Wiener-Hopf
factorization of $x_p$. The result is a special case of the
$\tau_2(p')$ free energy. For $r=N$, it and $L_r(t_p)$ can be 
obtained from (\ref{10d}). For $N=3$ we have verified that these 
results are consistent with previously obtained series expansions 
(eqns. 48 -- 52 of Ref. \cite{RJB98b}).

\end{document}